\documentclass[aps,prb,showpacs,twocolumn,superscriptaddress]{revtex4-1}

\usepackage{graphicx,amsmath,amssymb}
\usepackage[usenames]{color}
\usepackage{float}
\usepackage[T1]{fontenc}

\usepackage[colorlinks=true, citecolor=blue, urlcolor=blue, linkcolor=blue ]{hyperref}
\hypersetup{breaklinks=true}
\begin{document}

\bibliographystyle{apsrev4-1}

\title{Notes on Quantum oscillation for Hatsugai-Kohmoto model}
\author{Yin Zhong}
\email{zhongy@lzu.edu.cn}
\affiliation{School of Physical Science and Technology $\&$ Key Laboratory for
Magnetism and Magnetic Materials of the MoE, Lanzhou University, Lanzhou 730000, China}
\affiliation{Lanzhou Center for Theoretical Physics, Key Laboratory of Theoretical Physics of Gansu Province}
\begin{abstract}
Motivated by the non-Fermi liquid (NFL) phase in solvable Hatsugai-Kohmoto (HK) model and ubiquitous quantum oscillation (QO) phenomena observed in strongly correlated electron systems, e.g. cuprate high-Tc superconductor and topological Kondo insulator SmB$_{6}$, we have studied the QO in HK model in terms of a combination of analytical and numerical calculation. In the continuum limit, the analytical results indicate the existence of QO in NFL state and its properties can be described by Lifshitz-Kosevich-like formula. Furthermore, numerical calculations with Luttinger's approximation on magnetic-field-dependent density of state, magnetization and particle's density agree with the findings of analytical treatment. Although numerical simulation from exact diagonalization exhibits certain oscillation behavior, it is hard to extract its oscillation period and amplitude. Therefore, more work (particularly the large-scale numerical simulation) on this interesting issue is highly desirable and we expect the current study on HK model will be helpful to understand generic QO in correlated electron materials.
\end{abstract}

\maketitle
\section{Introduction}\label{intr}
Many-body physics beginning with the exploration of interacting electron gas, has been an essential issue in modern condensed matter physics, particularly after the discovery of cuprate high-Tc superconductors and fractional quantum Hall effect.\cite{Coleman2015} Recently, many researchers have focused on the so-called Hatsugai-Kohmoto (HK) model,\cite{Hatsugai1992,Baskaran1991,Hatsugai1996} which acts as a unique lattice fermion model since its solvability results from infinite-ranged interaction and it behaves as a Hubbard atom in momentum space. Motivated by above solvability in any spatial dimensional and electron filling, interesting and intriguing extensions of HK model have been invented and studied, including unconventional electron pairing, Fermi arc, Kondo impurity, many-band system and so on.\cite{Phillips2018,Yeo2019,Phillips2020,Yang2021,Zhu2021,Zhao2022,Setty2021,Mai2022,Huang2022,Li2022,Setty2020,Setty2021b,Zhong2022,Wang2023,Souza2023}
Unlike other well-established solvable models in recent decades, the HK-like models do not need any quenched disorder as in Sachdev-Ye-Kitaev model or local $Z_{2}$ gauge symmetry in Kitaev's toric code and honeycomb model.\cite{Sachdev,Maldacena,Chowdhury,Kitaev1,Kitaev2,Prosko,Zhong2013,Smith2017}

\begin{figure}
\includegraphics[width=0.75\linewidth]{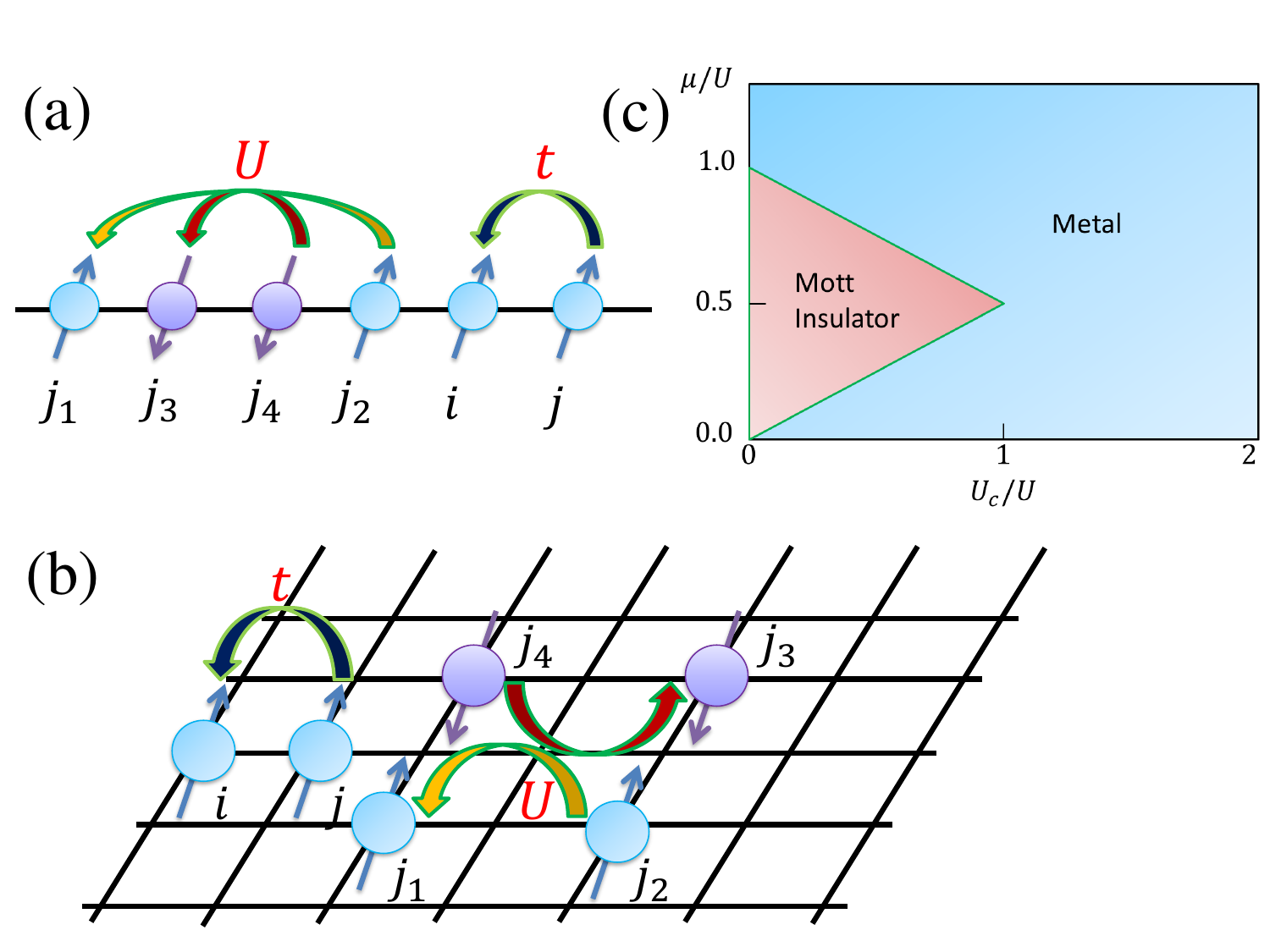}
\caption{\label{fig:fig_0} (a) The Hatsugai-Kohmoto (HK) model with hopping strength $t$ and interaction $U$ in one spatial dimension and (b) on a square lattice. (c) The exact ground-state phase diagram for HK model exhibits a
Mott insulator and a non-Fermi-liquid-like metal. ($U_{c}=W$ with $W$ being the band-width) The transition from metallic state to
gapped Mott insulating phase belongs to the universality of the continuous Lifshitz transition.}
\end{figure}

For the original HK model, it has translation invariance with topologically trivial nature, but surprisingly, it provides a strictly exact playground for non-Fermi liquid (NFL) and featureless Mott insulator in any spatial dimension, which is rare in statistical mechanics and condensed matter physics. (See Fig.~\ref{fig:fig_0}) The solvability of HK model results from its locality in momentum space and one can diagonalize HK Hamiltonian (just diagonal $4\times4$-matrix) for each momentum. The current studies have mainly focused on an interesting extension of HK model, i.e. the superconducting instability from the intrinsic NFL state in HK model,\cite{Phillips2020} which is inspired by ubiquitous NFL behaviors and their link to unconventional superconductivity in cuprate, iron-based superconductors (SC) and many heavy fermion compounds. Unexpected properties such as topological $s$-wave pairing and two-stage superconductivity have been discovered.\cite{Zhu2021,Zhao2022,Li2022}

At the same time, it is well-known that physical observables like magnetization and resistivity in metals show periodic oscillation under external magnetic field.\cite{Shoenberg1984} (de Haas-van Alphen and Shubnikov-de Haas effect) Such phenomena is called quantum oscillation (QO) and is generally believed to result from the oscillation of electron's density of state at Fermi energy when Landau level crosses Fermi energy periodically. The underlying microscopic description is captured by the famous Lifshitz-Kosevich (LK) formula, which provides a practical tool for metallic systems to extract electrons' effective mass and Fermi surface from period and amplitude of oscillation. Furthermore, validity of LK formula has been intensively examined and it is still valid if electron-electron interaction effect does not alter the Fermi liquid nature of the considered system.\cite{Luttinger1961,Wasserman1996} For practical calculations, the extended LK formula with electron's self-energy is widely used and it is found that several NFL-like self-energy still leads to LK-like results.\cite{Wasserman1996}

However, a natural question arises when the system is not described by Fermi liquid theory and the concept of quasiparticle breaks down. Thus, we may ask how about is the QO relating to NFL phenomena and the fate of LK theory.\cite{Chakravarty2011} This question is partially motivated by recent experiments of QO in many strongly correlated electron systems, including cuprate and iron-based superconductor,\cite{Sebastian2015,Carrington2011} heavy fermion compounds,\cite{McCollam2005} topological Kondo insulator,\cite{Tan2015} excitonic insulator and twisted-bilayer graphene.\cite{Wang2021,Cao2016} It is noted that in spite of the unambiguous NFL behaviors in cuprate and heavy fermion compound, or even insulating nature in topological Kondo insulator and excitonic insulator, QO has been firmly established in above quantum materials and certain tentative theoretical explanations have been put forward.\cite{Allais2014,Harrison2009,Knolle2015} In literature, electron correlation effect in QO has been widely studied by Hartree-Fock mean-field approximation,\cite{Doh1998,Allocca2022,Zyuzin2023} Hubbard-I approximation,\cite{Sherman2015} string-theory-inspired holographic duality,\cite{Denef2009,Hartnoll2010} slave-particle theory with large-N approximation and dynamic mean-field theory.\cite{Rasul1989,Galitski2009,Fritz2010,Acheche2017,Vucicevic2021} It should be emphasized that all these mentioned treatments involve either uncontrollable approximation like Hartree-Fock mean-field theory, artificial large-N limit or infinite-dimension limit, thus our understanding on QO in interacting many-electron system is obviously not complete.

In this work, inspired by the NFL state in solvable HK model, we study its possible QO. With the Luttinger's approximation, Hofstadter butterfly exists in all phases of the ground-state whatever they are NFL or Mott insulator. By examining the magnetic-field-dependent density of state, magnetization and particle's density, we find NFL states indeed show QO and their zero-temperature behaviors are captured by LK-like formula, which reflects the existence of two-Fermi-surface structure of the (non-Landau) quasiparticle in NFL. We have also performed an exact diagonalization calculation to go beyond the Luttinger's approximation. Due to the limited size of the system, although numerical results exhibit certain oscillation behavior, it is hard to extract its oscillation period and amplitude. Therefore, more work (particularly the large-scale numerical simulation) on this interesting issue is highly desirable and we expect the current study on HK model will be helpful to understand generic QO in strongly correlated electron materials.

The remaining parts of this paper are organized as follows. In Sec.~\ref{sec1}, the HK model will be introduced with a quick review about its basic properties. In Sec.~\ref{sec2}, with Luttinger's approximation, we calculate and discuss QO in HK model. Then, Sec.~\ref{sec3} provides an alternative calculation based on exact diagonalization. Sec.~\ref{sec4} is devoted to some discussions. Finally, we summarize our work in Sec.~\ref{sec5}.
\section{The Hatsugai-Kohmoto model}\label{sec1}
\subsection{Quick review of Hatsugai-Kohmoto model}
The Hatsugai-Kohmoto model we will study in this work has the following Hamiltonian,
\begin{eqnarray}
\hat{H}&=&-\sum_{i,j,\sigma}t_{ij}\hat{c}_{i\sigma}^{\dag}\hat{c}_{j\sigma}-\mu\sum_{j\sigma}\hat{c}_{j\sigma}^{\dag}\hat{c}_{j\sigma}\nonumber\\
&+&\frac{U}{N_{s}}\sum_{j_{1},j_{2},j_{3},j_{4}}\delta_{j_{1}+j_{3}=j_{2}+j_{4}}
\hat{c}_{j_{1}\uparrow}^{\dag}\hat{c}_{j_{2}\uparrow}\hat{c}_{j_{3}\downarrow}^{\dag}\hat{c}_{j_{4}\downarrow}\label{eq1}.
\end{eqnarray}
Here, the above model is defined on certain lattice, such as a one-dimensional chain or square lattice in Fig.~\ref{fig:fig_0}(a) and (b). Other lattices like triangular or Kagome lattice can also be considered, however such multi-sublattice structure may lead to extra complexity and the above model must to be modified.\cite{Mai2022}

In Eq.~\ref{eq1}, we use $\hat{c}_{j\sigma}^{\dag}$ to denote the creation operator of conduction electron ($c$-electron) at site $j$ with spin flavor $\sigma=\uparrow,\downarrow$. It satisfies the standard fermionic anti-commutation rule $\{\hat{c}_{i\sigma},\hat{c}_{j\sigma'}^{\dag}\}=\delta_{ij}\delta_{\sigma\sigma'}$.

Next, $t_{ij}$ is hopping integral between $i,j$ sites. Furthermore, to fix the electron's density, the chemical potential $\mu$ is added. $N_{s}$ is the number of sites. The last term of $\hat{H}$ is the HK interaction,\cite{Hatsugai1992} which is an infinite-ranged interaction between four electrons but preserves the center of motion for $c$-electron due to the constraint of $\delta$ function.

Most importantly, if we use the Fourier transformation $\hat{c}_{j\sigma}=\frac{1}{\sqrt{N_{s}}}\sum_{k}e^{ikR_{j}}\hat{c}_{k\sigma}$, it is found that the Hamiltonian is local in momentum space, which means,
\begin{eqnarray}
\hat{H}&=&\sum_{k}\hat{H}_{k}\nonumber\\
\hat{H}_{k}&=&\sum_{\sigma}(\varepsilon_{k}-\mu)\hat{c}_{k\sigma}^{\dag}\hat{c}_{k\sigma}+U
\hat{c}_{k\uparrow}^{\dag}\hat{c}_{k\uparrow}\hat{c}_{k\downarrow}^{\dag}\hat{c}_{k\downarrow}\label{eq2},
\end{eqnarray}
where $\varepsilon_{k}$ are dispersion of electrons and can be found from Fourier transformation of $t_{ij}$. In some sense, the interaction term $U
\hat{c}_{k\uparrow}^{\dag}\hat{c}_{k\uparrow}\hat{c}_{k\downarrow}^{\dag}\hat{c}_{k\downarrow}$ is just as the Hubbard interaction in momentum space and it may stabilize a non-trivial NFL fixed point due to the breaking of a hidden $Z_{2}$-symmetry.\cite{Huang2022} However, it seems that the true nature of such symmetry-breaking is still unclear.

Let us return to the discussion of Hamiltonian Eq.~\ref{eq2}. If we choose Fock state
\begin{equation}
|n_{1},n_{2}\rangle\equiv
(\hat{c}_{k\uparrow}^{\dag})^{n_{1}}|0\rangle(\hat{c}_{k\downarrow}^{\dag})^{n_{2}}|0\rangle\label{eq3}
\end{equation}
with $n_{i}=0,1$ as basis, $\hat{H}_{k}$ can be written as a diagonal $4\times4$ matrix, whose eigen-energy is $0,\varepsilon_{k}-\mu,\varepsilon_{k}-\mu,2(\varepsilon_{k}-\mu)+U$ and the corresponding eigen-state is $|0\rangle_{k}\equiv|00\rangle,|\sigma=\uparrow\rangle_{k}\equiv|10\rangle,|\sigma=\downarrow\rangle_{k}\equiv|01\rangle,|\uparrow\downarrow\rangle_{k}\equiv|11\rangle$, which means states are empty, single occupation with spin-up and spin-down, and double occupation.

Therefore, the many-body ground-state of $\hat{H}$ is just the direct-product state of each $\hat{H}_{k}$'s ground-state, i.e. $|\Psi_{g}\rangle=\prod_{k\in\Omega_{0}}|0\rangle_{k}\prod_{k\in\Omega_{1}}|\sigma\rangle_{k}\prod_{k\in\Omega_{2}}|\uparrow\downarrow\rangle_{k}$.($\Omega_{0},\Omega_{1},\Omega_{2}$ are the momentum range for different occupation) If $\Omega_{0}=\Omega_{2}=0$, each (momentum) state is only occupied by one electron, the system is a Mott insulator. Otherwise, we obtain a metallic state with non-Fermi liquid properties, e.g. the appearance of Luttinger surface, violation of Luttinger theorem,\cite{Phillips2020} non-Pauli spin susceptibility and exclusion statistics.\cite{Hatsugai1996,Vitoriano2000} Similarly, excited states and their energy are easy to be constructed, so $\hat{H}$ (Eq.~\ref{eq1}) has been solved since all eigen-states and eigen-energy are found.

For our purpose, it is useful to present the single-particle Green's function and some ground-state or thermodynamic quantities for HK model. For example, the single-particle Green's function can be obtained in terms of equation of motion, which reads as (See Appendix.\ref{ap_A})
\begin{eqnarray}
G_{\sigma}(k,\omega)&=&\frac{1+\frac{U\langle\hat{n}_{k\bar{\sigma}}\rangle}{\omega-(\varepsilon_{k}-\mu+U)}}{\omega-(\varepsilon_{k}-\mu)}\nonumber\\
&=&\frac{1-\langle\hat{n}_{k\bar{\sigma}}\rangle}{\omega-(\varepsilon_{k}-\mu)}+\frac{\langle\hat{n}_{k\bar{\sigma}}\rangle}{\omega-(\varepsilon_{k}-\mu+U)}\label{eq4}
\end{eqnarray}
where $\langle\hat{n}_{k\bar{\sigma}}\rangle$ is the expectation value of electron number operator $\hat{n}_{k\bar{\sigma}}=\hat{c}^{\dag}_{k\bar{\sigma}}\hat{c}_{k\bar{\sigma}}$ with spin $\bar{\sigma}=-\sigma$. For paramagnetic solution, one finds $\langle\hat{n}_{k\bar{\sigma}}\rangle=\frac{f_{F}(\varepsilon_{k}-\mu)}{f_{F}(\varepsilon_{k}-\mu)+1-f_{F}(\varepsilon_{k}-\mu+U)}$ with $f_{F}(x)=1/(e^{x/T}+1)$ being the Fermi distribution function. The pole of the above Green's function tells us that there exist two kinds of quasiparticle, i.e. holon $\hat{h}_{k\sigma}^{\dag}=\hat{c}_{k\sigma}(1-\hat{c}_{k\bar{\sigma}}^{\dag}\hat{c}_{k\bar{\sigma}})$ and doublon $\hat{d}_{k\sigma}^{\dag}=\hat{c}_{k\sigma}^{\dag}\hat{c}_{k\bar{\sigma}}^{\dag}\hat{c}_{k\bar{\sigma}}$. To see their significance, one finds that
$\hat{h}_{k\uparrow}^{\dag}|10\rangle=|00\rangle$, $\hat{h}_{k\uparrow}^{\dag}|00\rangle=\hat{h}_{k\uparrow}^{\dag}|01\rangle=\hat{h}_{k\uparrow}^{\dag}|11\rangle=0$, which means $\hat{h}^{\dag}_{k\uparrow}$ creates the state with no electron, i.e. a hole state. Similarly, $\hat{d}_{k\uparrow}^{\dag}|01\rangle=|11\rangle$, $\hat{d}_{k\uparrow}^{\dag}|00\rangle=\hat{h}_{k\uparrow}^{\dag}|10\rangle=\hat{h}_{k\uparrow}^{\dag}|11\rangle=0$ and it states that $\hat{d}_{k\uparrow}^{\dag}$ creates the state with two occupied electrons. The corresponding quasiparticle energy bands are $\varepsilon_{k}-\mu,\varepsilon_{k}-\mu+U$. We should emphasize that they are not Landau quasiparticle since adiabatical continuity into non-interacting limit $U=0$ does not work.

Next, at finite-$T$, the thermodynamics of HK model is determined by
its free energy density $f$, which is related to partition function $\mathcal{Z}$ as
\begin{eqnarray}
f=-\frac{T}{N_{s}}\ln\mathcal{Z},\mathcal{Z}=\mathrm{Tr }e^{-\beta \hat{H}}=\prod_{k}\mathrm{Tr} e^{-\beta \hat{H}_{k}}=\prod_{k}f_{k}\label{eq5}
\end{eqnarray}
Here, one notes that the partition function is easy to calculate since each $k$-state contributes independently. We have defined $f_{k}=1+2z_{k}+z_{k}^{2}e^{-\beta U}$ and $z_{k}=e^{-\beta(\varepsilon_{k}-\mu)}$. At zero temperature, the free energy density reduces into the ground-state energy density, which has very simple expression,
\begin{equation}
e_{g}=\frac{1}{N_{s}}\sum_{k}[(\varepsilon_{k}-\mu)\theta(\mu-\varepsilon_{k})+(\varepsilon_{k}-\mu+U)\theta(\mu-\varepsilon_{k}-U)],\label{eq6}
\end{equation}
where $\theta(x)$ is the standard unit-step function ($\theta(x)=1$ for $x>0$ and $\theta(x)=0$ if $x<0$). Therefore, the electron density at $T=0$ is found to be
\begin{equation}
n=\frac{1}{N_{s}}\sum_{k}[\theta(\mu-\varepsilon_{k})+\theta(\mu-\varepsilon_{k}-U)].\label{eq7}
\end{equation}
which indicates two Fermi surfaces located at $\varepsilon_{k}=\mu$ and $\varepsilon_{k}=\mu-U$.
\subsection{Adding orbital magnetic field}
Because we are interested in possible QO of HK model, the orbital magnetic field should be included. This can be realized easily via Pierls substitution as $t_{ij}\rightarrow t_{ij}e^{ia_{ij}}$, where phase factor $a_{ij}=\frac{e}{\hbar}\int_{r_{j}}^{r_{i}}d\vec{r}'\cdot \vec{A}(\vec{r}')$ and $\vec{A}$ is the vector potential of external magnetic field. Then, we can write HK model as $\hat{H}=\hat{H}_{0}+\hat{H}_{\mu}+\hat{H}_{U}$
\begin{eqnarray}
&&\hat{H}_{0}=-\sum_{i,j,\sigma}t_{ij}e^{ia_{ij}}\hat{c}_{i\sigma}^{\dag}\hat{c}_{j\sigma},\nonumber\\
&&\hat{H}_{\mu}=-\mu\sum_{j\sigma}\hat{c}_{j\sigma}^{\dag}\hat{c}_{j\sigma},~~~~\nonumber\\
&&\hat{H}_{U}=\frac{U}{N_{s}}\sum_{j_{1},j_{2},j_{3},j_{4}}\delta_{j_{1}+j_{3}=j_{2}+j_{4}}
\hat{c}_{j_{1}\uparrow}^{\dag}\hat{c}_{j_{2}\uparrow}\hat{c}_{j_{3}\downarrow}^{\dag}\hat{c}_{j_{4}\downarrow}.\label{eq8}
\end{eqnarray}
Frankly speaking, the above Hamiltonian is hard to solve since adding external magnetic field breaks the translation invariance, which leads to the violation of solvability of the original HK model (Eq.~\ref{eq1}).

Instead, following the original treatment of Luttinger,\cite{Luttinger1961,Wasserman1996} whose approximation is able to capture QO in interacting electron systems such as Fermi liquid, we can still use results of last section but only need to replace single-particle energy $\varepsilon_{k}$ with eigen-energy of $\hat{H}_{0}$ (denoted as $\{E_{m}\}$). In the language of many-body physics, such replacement is a statement that the external magnetic field only modifies the single-particle spectrum but not the expression of self-energy function.

Therefore, the ground-state energy density under external magnetic field ($\vec{B}=\nabla\times\vec{A}$) has the new formalism,
\begin{equation}
e_{g}=\frac{1}{N_{s}}\sum_{m}[(E_{m}-\mu)\theta(\mu-E_{m})+(E_{m}-\mu+U)\theta(\mu-E_{m}-U)],\label{eq9}
\end{equation}
and the related electron density is
\begin{equation}
n=\frac{1}{N_{s}}\sum_{m}[\theta(\mu-E_{m})+\theta(\mu-E_{m}-U)].\label{eq10}
\end{equation}
Thus, the magnetization and susceptibility in ground-state can be found by $M=-\frac{\partial e_{g}}{\partial B},\chi_{s}=\frac{\partial M}{\partial B}$. For thermodynamics at finite $T$, the only modification is $f_{k}\rightarrow f_{m}=1+2z_{m}+z_{m}^{2}e^{-\beta U}$ with $z_{m}=e^{-\beta(\varepsilon_{m}-\mu)}$ in Eq.~\ref{eq5}.

To be specific, let us consider the square lattice on $xy$-plane with nearest-neighbor-hopping ($t_{ij}\rightarrow t$) and the magnetic field is along the $z$-axis $\vec{B}=B\vec{e}_{z}$. In Landau gauge, the corresponding vector potential is $\vec{A}=Bx \vec{e}_{y}$, so $a_{i,i+x}=0$ and $a_{i,i+y}=\frac{e}{\hbar}Ba^{2}i_{x}$, ($a$ is the lattice constant and will be set to unit) which gives rise to
\begin{eqnarray}
\hat{H}_{0}=-t\sum_{i,\sigma}(\hat{c}_{i\sigma}^{\dag}\hat{c}_{i+x,\sigma}+e^{i\frac{e}{\hbar}Bi_{x}}\hat{c}_{i\sigma}^{\dag}\hat{c}_{i+y,\sigma}).\label{eq11}
\end{eqnarray}
and the phase factor part can be parameterized as $e^{i\frac{e}{\hbar}Bi_{x}}=e^{i\frac{e}{\hbar}\Phi i_{x}}=e^{i2\pi w i_{x}}$ with flux $\Phi=2\pi\frac{\hbar}{e}w$ on each plaquette. In fact $\hat{H}_{0}$ is just the famous Hofstadter model with spin degree of freedom and has been widely studied in recent years due to its realization in cold atom experiments and its relevance to moir\'{e} superlattice systems.\cite{Hofstadter1976,Aidelsburger2013,Miyake2013,Kennedy2015,Dean2013,Hunt2013,Spanton2018}

After diagonalization of $\hat{H}_{0}$, we obtain its eigen-energy $\{E_{m}\}$ for given magnetic field $B$, thus interesting quantities can be calculated straightforwardly.
\section{Quantum oscillation in non-Fermi liquid: The case study on Hatsugai-Kohmoto model}\label{sec2}
\subsection{Lattice calculation}
In this section, we study the HK model under magnetic field on square lattice whose single-particle Hamiltonian is $\hat{H}_{0}$ (Eq.~\ref{eq11}).

\begin{figure}
\includegraphics[width=0.75\linewidth]{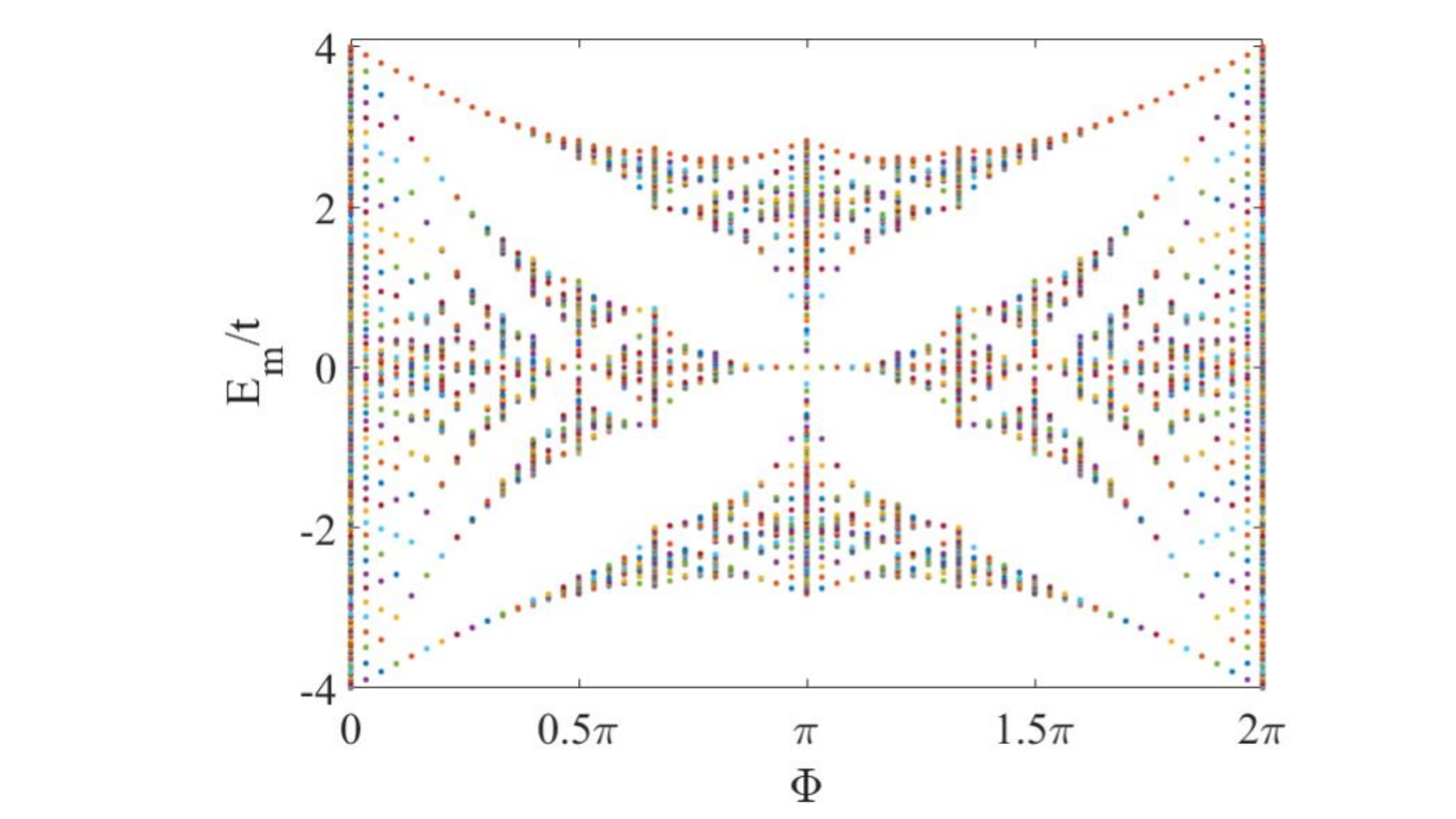}
\caption{\label{fig:1} The butterfly spectrum of $\hat{H}_{0}$ for $E_{m}$ versus flux $\Phi=2\pi w=B$.}
\end{figure}

Firstly, for parameters $t=e=\hbar=1$, the single-particle spectrum $\{E_{m}\}$ has been shown in Fig.~\ref{fig:1}, in which we consider a $60\times60$ system with periodic boundary condition. It is clear to see the well-known butterfly spectrum of non-interacting Hamiltonian $\hat{H}_{0}$ for $E_{m}$ versus flux $\Phi=2\pi w=B$.\cite{Hofstadter1976}

\subsubsection{Density of state of electron}
Now, we turn to the interacting case with the full Hamiltonian $\hat{H}=\hat{H}_{0}+\hat{H}_{\mu}+\hat{H}_{U}$. Since the concept of single-particle spectrum is meaningless in interacting case, we consider the density of state (DOS) of electron, whose expression is found to be
\begin{eqnarray}
N(\omega,\Phi)&=&\frac{1}{N_{s}}\sum_{m}[(1-n_{m})\delta(\omega-E_{m}+\mu)\nonumber\\
&+&n_{m}\delta(\omega-E_{m}+\mu-U)],
\end{eqnarray}
where $n_{m}=(\theta(\mu-E_{m})+\theta(\mu-E_{m}-U))/2$.
\begin{figure}
\flushleft
\includegraphics[width=1.25\linewidth]{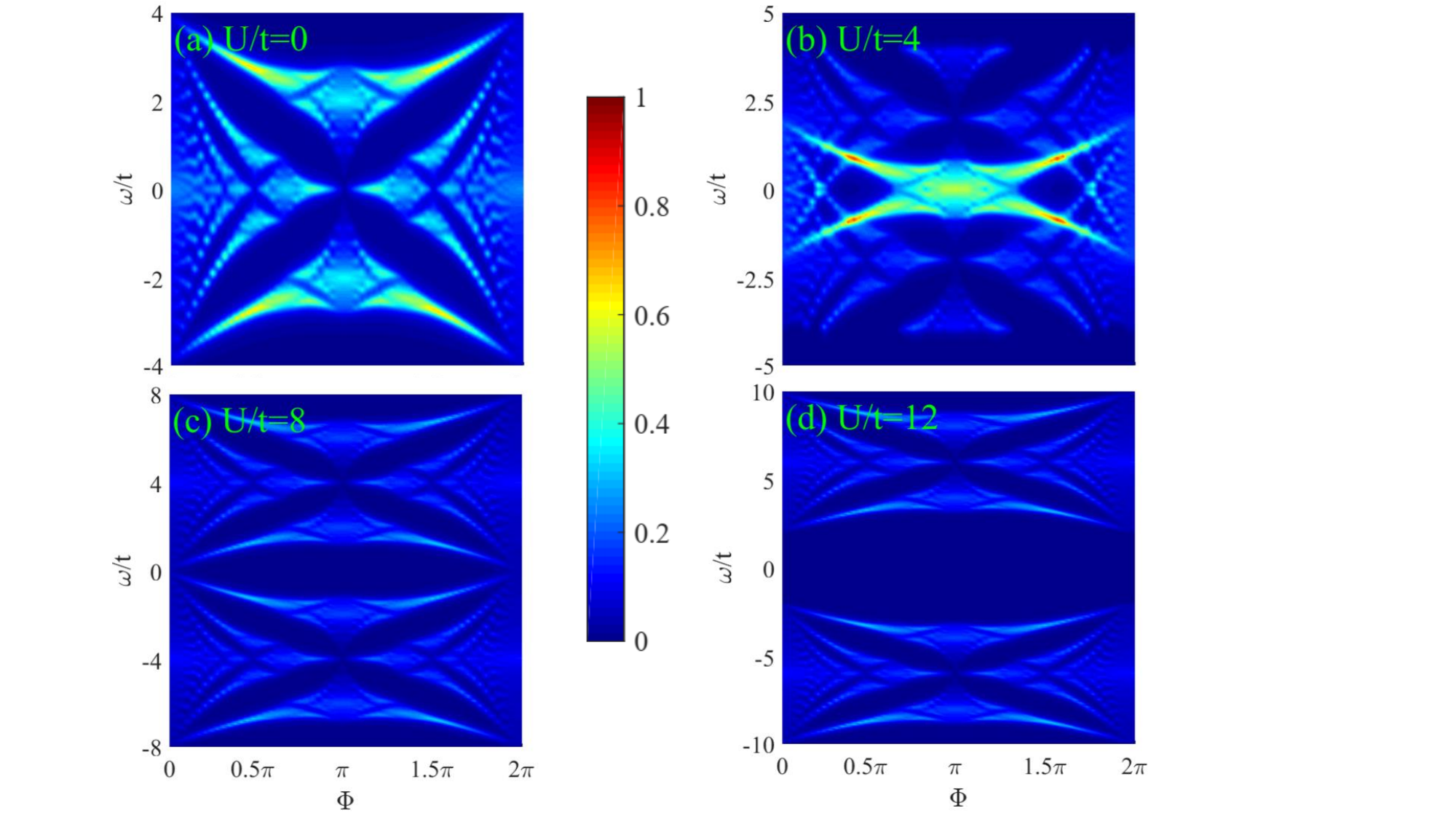}
\caption{\label{fig:5} The density of state of electron $N(\omega,\Phi)$ for different interaction $U$ at half-filling.}
\end{figure}

\begin{figure}
\flushleft
\includegraphics[width=1.1\linewidth]{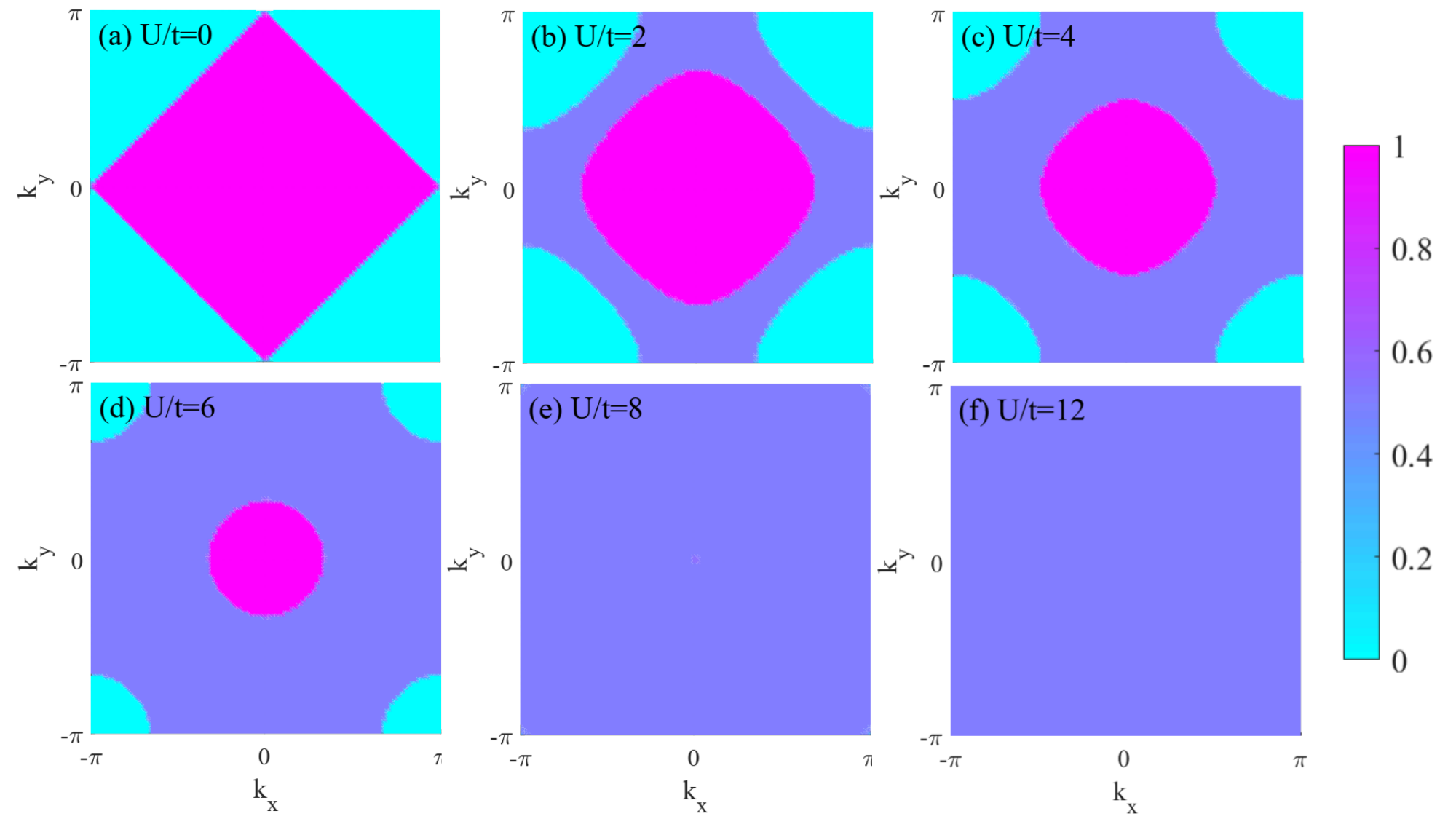}
\caption{\label{fig:6} The electron's distribution function $n(k)$ in momentum space for different interaction $U$ at half-filling.}
\end{figure}

In Fig.~\ref{fig:5}, we have plotted $N(\omega,\Phi)$ for $U/t=0,4,8,12$ at half-filling ($\mu=U/2$). These interaction parameters are chosen to represent typical regimes in the ground-state phase diagram without magnetic field, i.e. non-interacting regime ($U/t=0$), NFL metal regime ($U/t=4$), NFL-Mott insulator transition point $U/t=8$ and Mott insulating regime ($U/t=12$). For completeness, the electron's distribution function $n(k)$ in momentum space for different interaction $U$ at half-filling has also been shown in Fig.~\ref{fig:6}. One notes that at half-filling, the NFL phase always has two-Fermi-surface structure, which is embodied by the jumps of electron's occupation from $1$ to $0.5$ and from $0.5$ to $0$.

It is interesting to find that in all cases we have studied, the Hofstadter butterfly exists in spite of the increasing of interaction. A noticeable feature of interaction is that it can split one butterfly in $U/t=0$ into two identical pieces. If the HK interaction is larger than the band-width of non-interacting electron, (Fig.~\ref{fig:5}(d)) there exists sensible energy gap between two butterflies, which is the reminiscent of the lower and upper Hubbard bands without external magnetic field.
We note that these features are similar to the findings in Falicov-Kimball model above the charge-density-wave transition, where numerically exact Monte Carlo simulation can be performed.\cite{Wrobel2010}

We all know that the conventional wisdom of QO relies on the oscillation behavior of density of state of electron at Fermi energy ($N(0,\Phi)\equiv N(0)$). If $N(0)$ has sharp oscillation behavior versus magnetic field $\Phi$, then physical observables like magnetization $M$, specific heat $C_{V}$ and resistivity $\rho$ must show QO as well. (recall that in Fermi liquid or Fermi gas, we have $M\sim N(0), C_{V}\sim N(0)T$)

\begin{figure}
\flushleft
\includegraphics[width=1.1\linewidth]{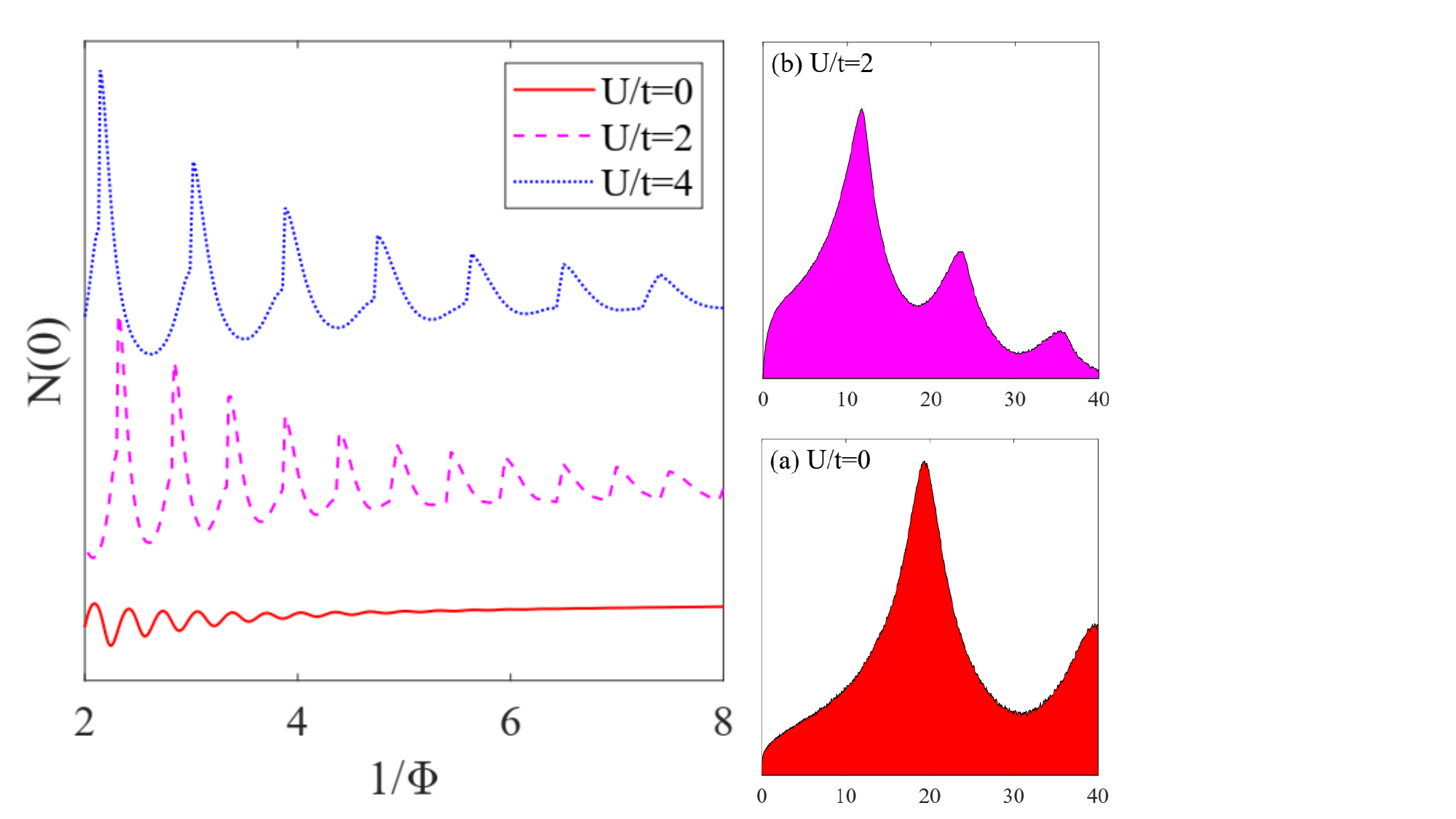}
\caption{\label{fig:7} The density of state of electron at Fermi energy $N(0)$ for different interaction $U$ at half-filling. (a)(b) Frequency of QO extracted from $N(0)$ for $U/t=0,2$.}
\end{figure}
Because Mott insulator has vanished $N(0)$, QO is not expected in this insulating phase. Therefore, we focus on metallic NFL state and in Fig.~\ref{fig:7}, it is clear to see QO of $N(0)$. (To have regular data on square lattice, we have used periodic boundary condition along $y$-direction while an open boundary condition is assumed for $x$-direction.) Performing (fast) Fourier transformation for $N(0)$, one is able to extract the period (or frequency) of QO, which relates to the area of the underlying Fermi surface. For example, Fig.~\ref{fig:7}(a) tells us that the frequency for non-interacting case is about $19.38$, which is quite near the exact Fermi surface area $A_{F}=(2\pi)^2/2=19.74$. (see also Fig.~\ref{fig:6}(a)) Furthermore, Fig.~\ref{fig:7}(b) gives two Fermi surface areas $A_{F1}=11.73$ and $A_{F2}=23.50$, which should be contrast with results $A_{F1}=11.99,A_{F2}=26.99$ from Fig.~\ref{fig:6}(b). Therefore, our analysis on oscillation of density of state indicates that the QO in NFL of HK model is rooted on the Fermi surface structure although the Landau quasiparticle disappears in this situation due to HK interaction.

\subsubsection{Magnetization and particle density}
\begin{figure}
\flushleft
\includegraphics[width=1.2\linewidth]{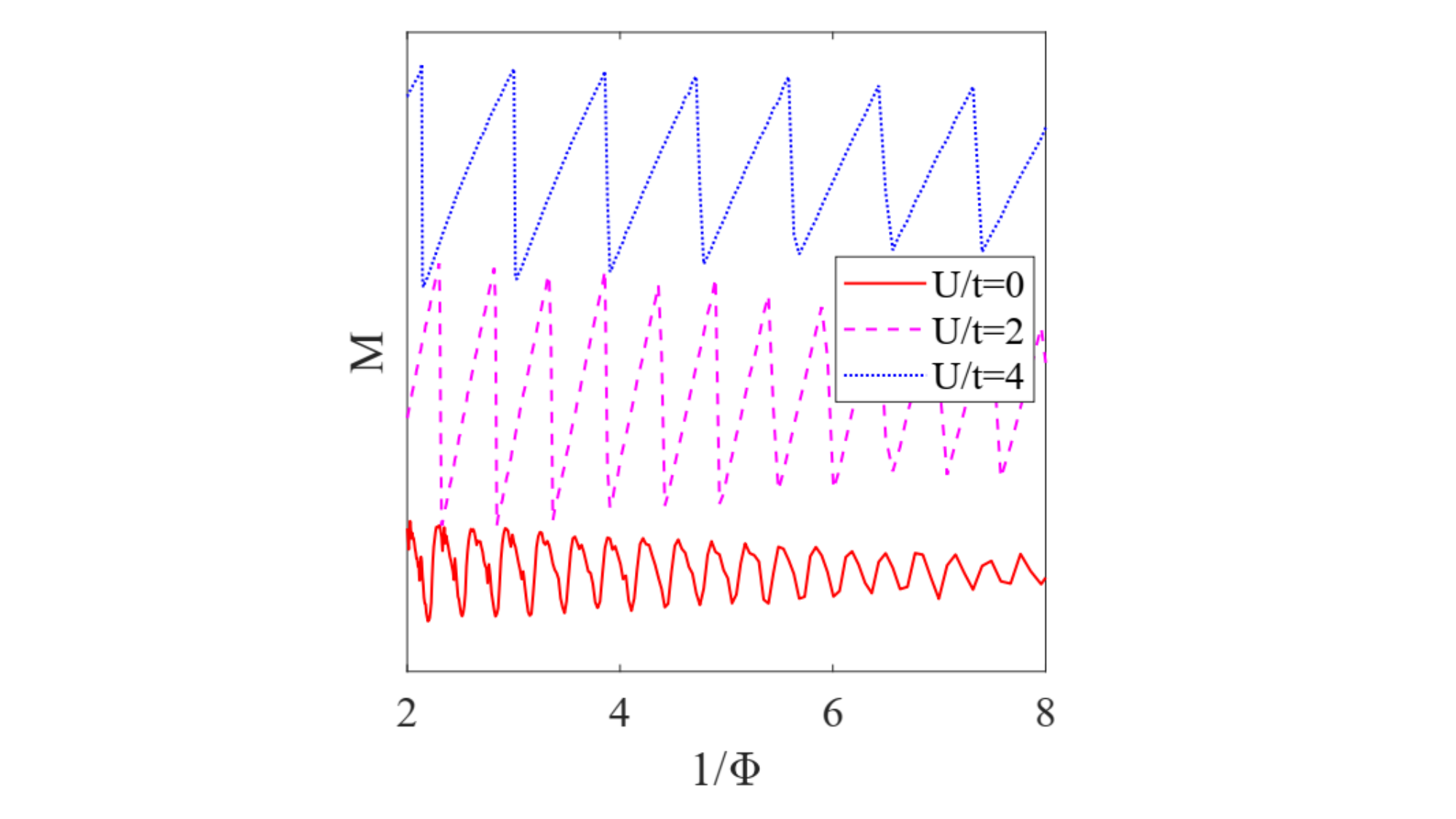}
\caption{\label{fig:8} The magnetization $M$ versus $1/\Phi$ for different interaction $U$.}
\end{figure}
Since the de Haas-van Alphen effect is encoded by studying the magnetization $M$ versus magnetic field, Fig.~\ref{fig:8} has given examples on $M$ for different $U$ in NFL state. Obviously, QO in $M$ is identified and as expected, it gives identical frequency of QO to $N(0)$. This fact reflects that the thermodynamics is indeed contributed from quasiparticles but not any mysterious quantum excitations like Majorana fermions proposed in topological Kondo insulator.\cite{Baskaran2015,Varma2020} Furthermore, a close look at data implies that the oscillation in $M$ is sharper and more regular than $N(0)$.

At the same time, using Eq.~\ref{eq10}, we have plotted particle density $n$ versus $\Phi$ and $1/\Phi$ for different interaction strength $U$.
Because of the particle-hole symmetry, $n$ at half-filling ($\mu=U/2$) will not show any oscillation behavior, therefore, we fix $\mu=0$ such that we are inspecting the system away from half-filling.
\begin{figure}
\includegraphics[width=1\linewidth]{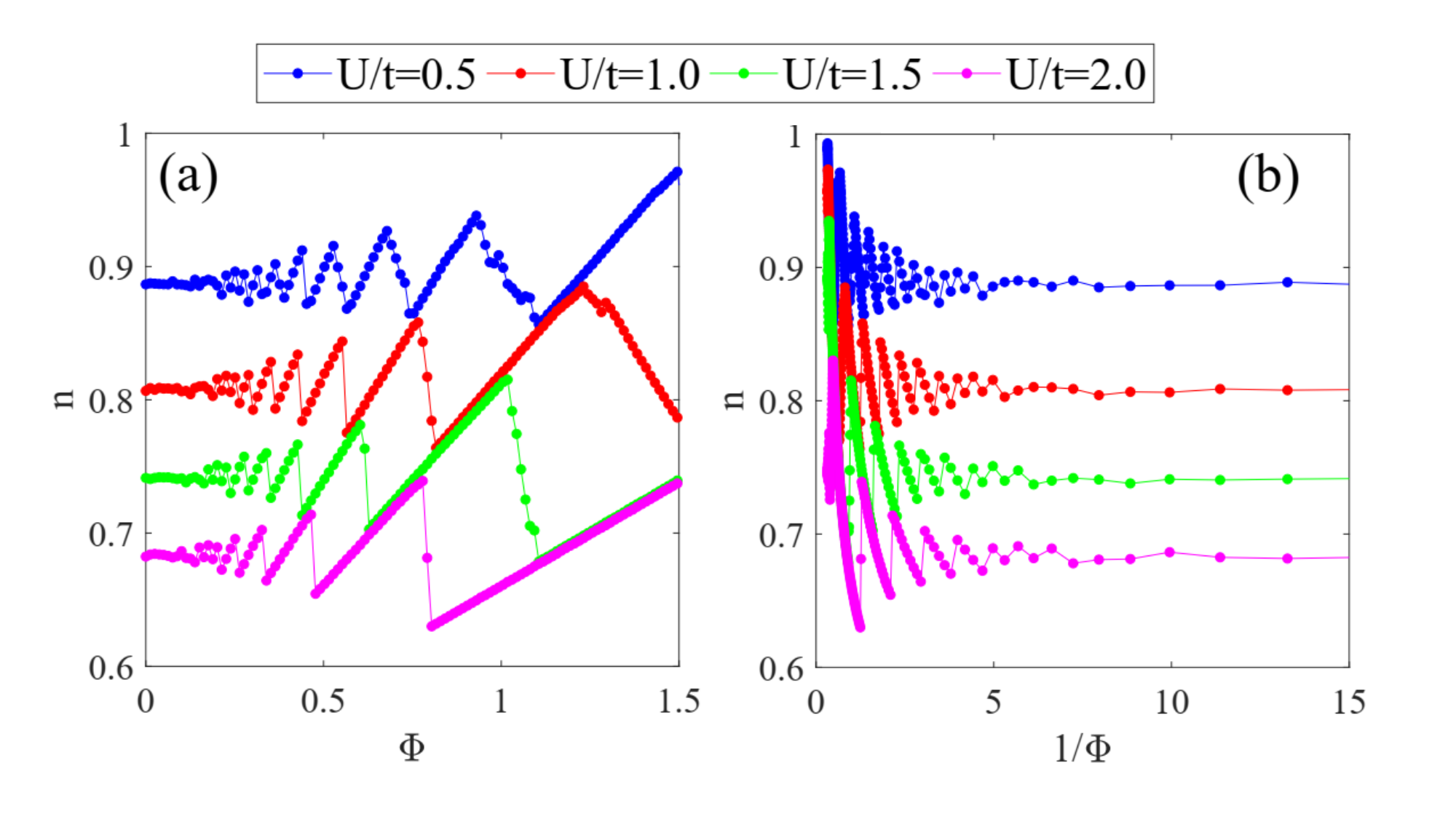}
\caption{\label{fig:2} The particle density $n$ versus $\Phi$ (a) and $1/\Phi$ (b) for different $U/t$.}
\end{figure}
It is not surprising that a clear signature of QO still exists in the particle density as shown in Fig.~\ref{fig:2}.

\subsection{Finite temperature effect}
\begin{figure}
\includegraphics[width=1\linewidth]{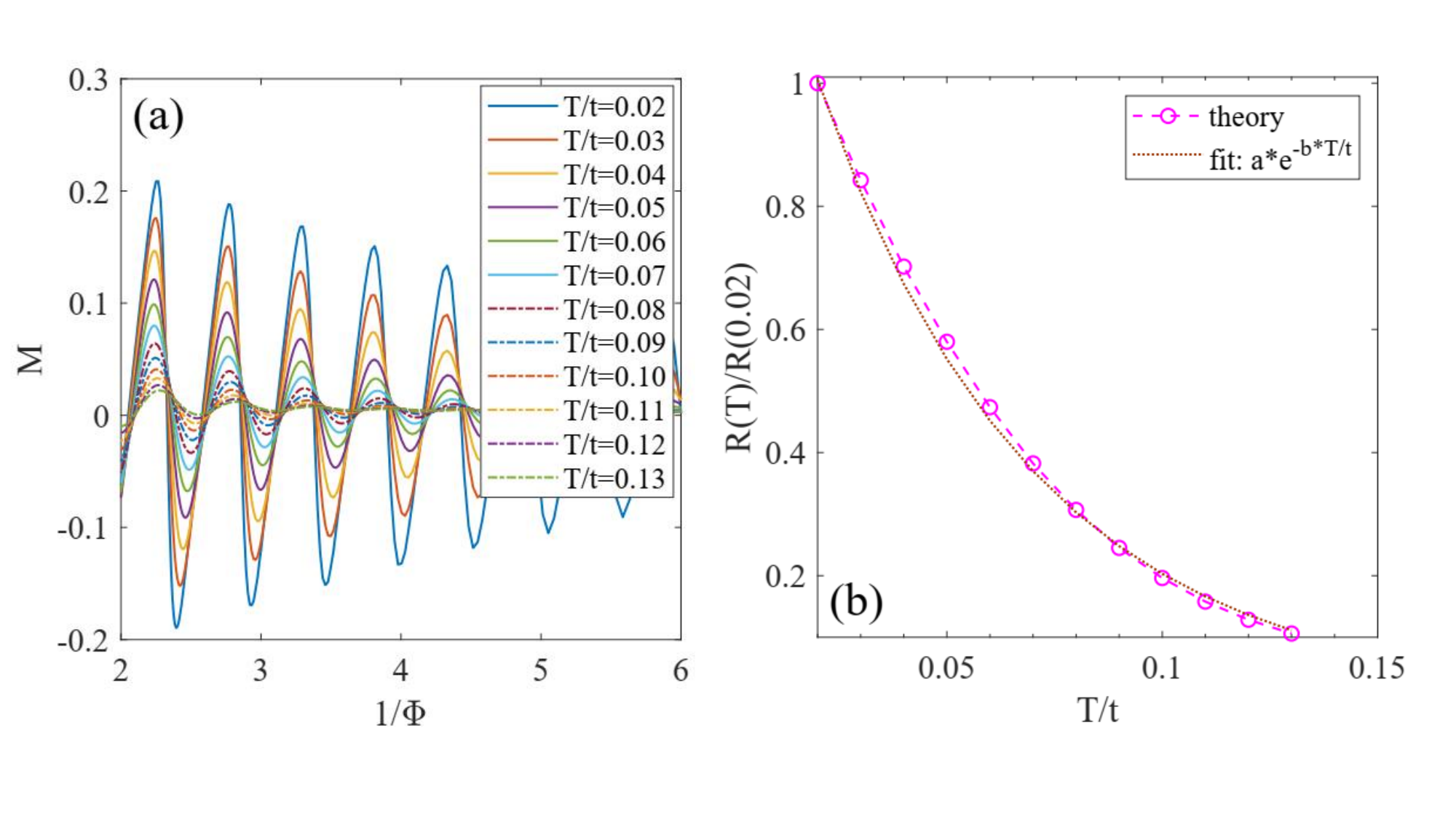}
\caption{\label{fig:9} (a) Magnetization $M$ versus $\Phi$ for different temperature. (b) The amplitude of QO ($R(T)$) versus temperature $T$. ($U/t=2,\mu=U/2$)}
\end{figure}

After presenting the results at $T=0$, here, we discuss the effect of temperature on QO. Specifically, we extract the amplitude of QO ($R(T)$) in NFL regime, which is a function of temperature. In Fig.~\ref{fig:9}, we have seen that the magnetization $M$ changes for different temperature $T$ and its amplitude decreases if $T$ increases. ($U/t=2,\mu=U/2$) Moreover, a fit with $a e^{-b T/t}$ works well if we try to fit the amplitude $R(T)$, which seems to agree with the well-established LK formula, i.e. $R_{LK}(T)\sim \frac{T}{\sinh T}\sim e^{-T}$.

\subsection{Continuum limit}
Although the direct lattice calculation provides the information of QO, in this subsection, we try to understand the origin of QO from the continuum limit.

In continuum limit, one can approximate electron's dispersion $\varepsilon_{k}$ with free particle energy $\frac{k^{2}}{2m_{e}}$. Meanwhile,  $\varepsilon_{m}=\hbar\omega_{c}(m+1/2)$ if we consider the usual Landau gauge. ($\omega_{c}=eB/m_{e}$) Then, the electron's density reads
\begin{equation}
n=\frac{1}{N_{\Lambda}}\sum_{k_{y}}\sum_{m=0}^{\Lambda}[\theta(\mu-\hbar\omega_{c}(m+1/2))+\theta(\mu-\hbar\omega_{c}(m+1/2)-U)].\label{eq12}
\end{equation}
Here, $\Lambda$ is the cutoff of Landau level index, which acts as a band-width. The factor $N_{\Lambda}$ has been introduced to give sensible definition of electron's density. The summation over $k_{y}$ gives the degeneracy of Landau level ($D$).

To proceed, we can rewrite $n$ as follows,
\begin{eqnarray}
n&=&\frac{D}{N_{\Lambda}}\lim_{\beta\rightarrow\infty}\int_{0}^{\infty}d\omega\frac{\sum_{m}\delta(\omega-\varepsilon_{m})+\sum_{m}\delta(\omega-\varepsilon_{m}-U)}{e^{\beta(\omega-\mu)}+1}\nonumber\\
&=&\frac{D}{N_{\Lambda}}\lim_{\beta\rightarrow\infty}\int_{0}^{\infty}d\omega\frac{1}{e^{\beta(\omega-\mu)}+1}\nonumber\\
&\times&\mathrm{Re}\int_{0}^{\infty}d\tau\sum_{m}(e^{i(\omega-\varepsilon_{m})\tau}
+e^{i(\omega-\varepsilon_{m}-U)\tau}).\label{eq13}
\end{eqnarray}
Then, using Poisson summation formula
$\sum_{m}e^{-im\hbar\omega_{c}\tau}=\frac{2\pi}{\hbar\omega_{c}}\sum_{p=0,\pm1,\pm2...}\delta(\tau-\frac{2\pi p}{\hbar\omega_{c}})$, and integrate over $\tau$, one finds
\begin{eqnarray}
n&=&\frac{2\pi D}{N_{\Lambda}\hbar\omega_{c}}\lim_{\beta\rightarrow\infty}\int_{0}^{\infty}d\omega\left(\frac{1}{e^{\beta(\omega-\mu)}+1}+\frac{1}{e^{\beta(\omega-\mu+U)}+1}\right)\nonumber\\
&\times&\mathrm{Re}\sum_{p}e^{i\frac{2\pi p}{\hbar\omega_{c}}(\omega-\frac{\hbar\omega_{c}}{2})}\nonumber\\
&\propto&\lim_{\beta\rightarrow\infty}\frac{2\pi}{\beta\hbar\omega_{c}}\mathrm{Re}\sum_{\omega_{n}>0}\sum_{p>0}(-1)^{p}\nonumber\\
&\times&\left(e^{i\frac{2\pi p}{\hbar\omega_{c}}(\mu+i\omega_{n})}+e^{i\frac{2\pi p}{\hbar\omega_{c}}(\mu-U+i\omega_{n})}\right)\nonumber\\
&=&\lim_{\beta\rightarrow\infty}\sum_{p>0}\frac{(-1)^{p}}{2\pi p}\frac{X}{\sinh X}\left(\cos\frac{2\pi p\mu}{\hbar\omega_{c}}+\cos\frac{2\pi p(\mu-U)}{\hbar\omega_{c}}\right).\nonumber\\
&&\label{eq14}
\end{eqnarray}
Here, $X=\frac{2\pi^{2}pT}{\hbar\omega_{c}}$ and one defines $R_{T}=\frac{X}{\sinh X}$ as damping factor due to temperature. It is amusing to note that the above Eq.~\ref{eq14} is just the well-known LK formula with oscillation frequency $\mu/(\hbar\omega_{c})$ and $(\mu-U)/(\hbar\omega_{c})$. When $T\rightarrow0$ or $\beta\rightarrow\infty$, we obtain $n\propto\sum_{p>0}\frac{(-1)^{p}}{2\pi p}\left(\cos\frac{2\pi p\mu}{\hbar\omega_{c}}+\cos\frac{2\pi p(\mu-U)}{\hbar\omega_{c}}\right)$. It is interesting to see that although the system is in NFL state and no Landau quasiparticle exits, the interaction does not lead to Dingle-like factor $e^{-\frac{U}{\hbar\omega_{c}}}$. This may be due to the coherent feature of non-Landau quasiparticles, i.e. doublon $\hat{d}_{k\sigma}$ and holon $\hat{h}_{k\sigma}$. Thus, a Boltzman transport theory for those non-Landau quasiparticles could be constructed if one can generalize the work on exclusion statistics.\cite{Bhaduri1996} We expect this theory will be useful to understand semiclassical transport of HK model.

Thus, it seems that under the approximation of Luttinger, despite the NFL nature, QO exists in HK model and it has two oscillation periods determined by chemical potential $\mu$ and $\mu-U$. We note that the finding of two oscillation periods is consistent with the Fourier analysis in previous subsection.

\section{Exact calculation based on exact diagonalization}\label{sec3}
Until now, our calculation is based on Luttinger's approximation, however, to our knowledge, the validity of such approximation for the HK model is not known.
The even worse situation may be that the basis of Luttinger's approximation, namely the Luttinger-Ward functional, is ill defined since its skeleton Feynman diagrams expansion breaks down.\cite{Kozik2015}

Instead, one may use (numerical) exact diagonalization to extract reliable information from our model (Eq.\ref{eq8} and \ref{eq11}). Since $y$-direction of our model has been chosen to have periodic boundary condition under Landau gauge, we can use Fourier transformation to write
\begin{equation}
\hat{c}_{j\sigma}=\frac{1}{\sqrt{L_{y}}}\sum_{k_{y}}e^{ik_{y}j_{y}}\hat{c}_{j_{x}\sigma}(k_{y}),\label{eq15}
\end{equation}
where $x-$direction remains to be intact. Then, the Hamiltonian reads as $\hat{H}=\sum_{k_{y}}\hat{H}(k_{y})$,
\begin{eqnarray}
\hat{H}(k_{y})&=&-t\sum_{j_{x}\sigma}(\hat{c}_{j_{x}\sigma}^{\dag}(k_{y})\hat{c}_{j_{x}+1\sigma}(k_{y})+\hat{c}_{j_{x}+1\sigma}^{\dag}(k_{y})\hat{c}_{j_{x}\sigma}(k_{y}))\nonumber\\
&+&\sum_{j_{x}\sigma}\left(-2t\cos(\Phi j_{x}+k_{y})-\mu\right)\hat{c}_{j_{x}\sigma}^{\dag}(k_{y})\hat{c}_{j_{x}\sigma}(k_{y})\nonumber\\
&+&\frac{U}{L_{x}}\sum_{j_{1x},j_{2x},j_{3x},j_{4x}}\delta_{j_{1x}+j_{3x}=j_{2x}+j_{4x}}
\hat{c}_{j_{1x}\uparrow}^{\dag}(k_{y})\hat{c}_{j_{2x}\uparrow}(k_{y})\nonumber\\
&\times&\hat{c}_{j_{3x}\downarrow}^{\dag}(k_{y})\hat{c}_{j_{4x}\downarrow}(k_{y}).\label{eq16}
\end{eqnarray}
The above Hamiltonian means that for given momentum $k_{y}$, one can just study the reduced object $\hat{H}(k_{y})$, which is an effective one-dimensional HK model with site-dependent potential $-2t\cos(\Phi j_{x}+k_{y})$. In principle, if one is interested in the exact thermodynamics or dynamics of HK model, each Hamiltonian $\hat{H}(k_{y})$ can be solved in terms of exact diagonalization up to $L_{x}$ sites along $x-$direction (i.e. $j_{x}=1,2...L_{x}$). Unfortunately, the largest system size which can be diagonalized is about $L_{x}^{max}=16$ and the infinite-ranged HK interaction must reduce the value of $L_{x}^{max}$.

\begin{figure}
\includegraphics[width=0.80\linewidth]{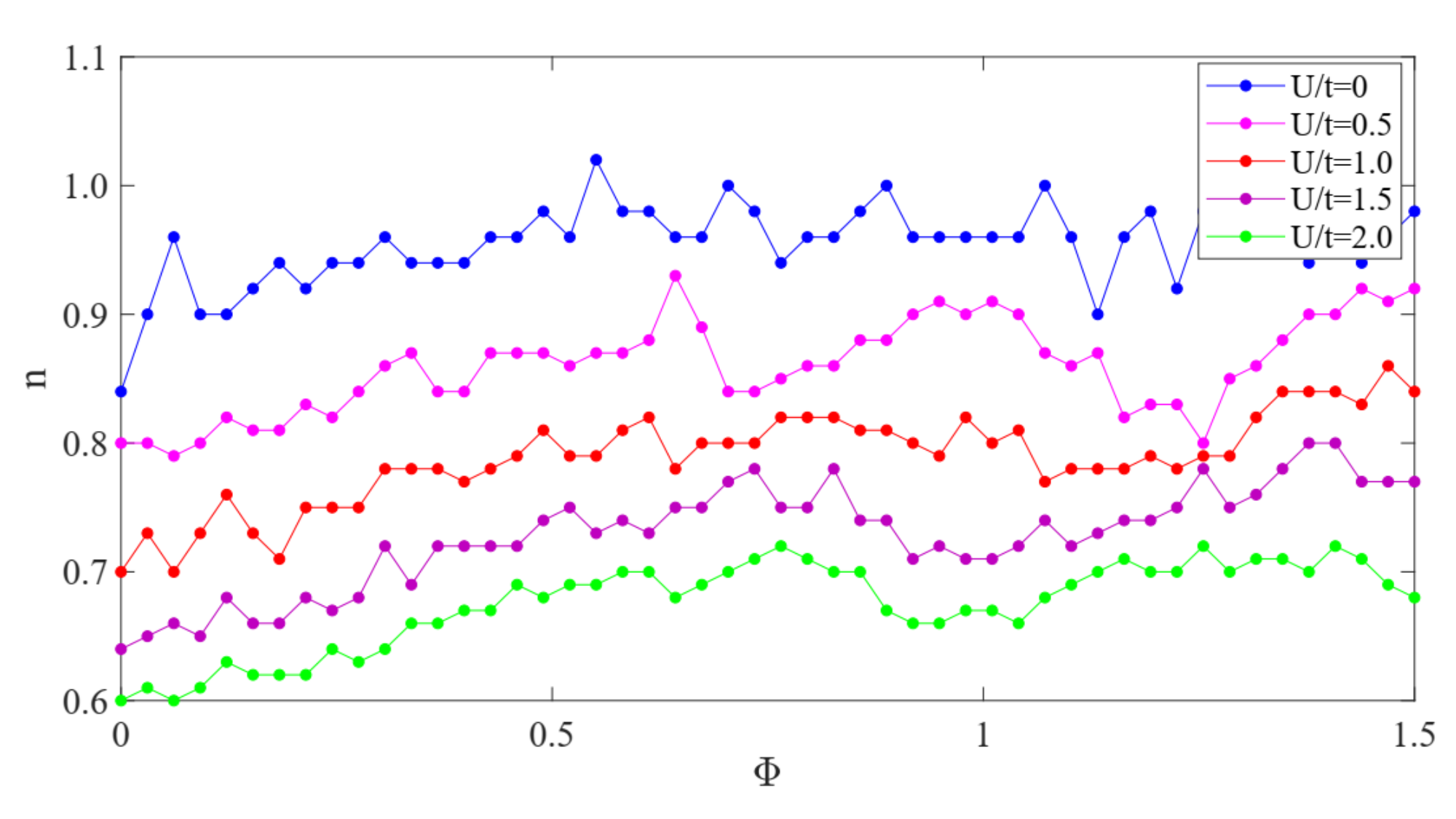}
\caption{\label{fig:4} The particle density $n$ versus $\Phi$ for $U/t=0,0.5,1.0,1.5,2.0$ from exact diagonalization.}
\end{figure}

In Fig.~\ref{fig:4}, we have shown the particle density $n$ versus $\Phi$ for $L_{x}=6$ and $L_{y}=16$. However, due to the limited size, particularly the smallness of $L_{x}$, although certain oscillation behavior appears, it is hard to detect clear QO and extract its oscillation period and amplitude. Therefore, much larger size of $L_{x}$ should be used to
explore the true signature of QO in HK model. In this direction, matrix-product-state or density-matrix-renormalization-group algorithm is the method of choice,\cite{Schollwock2005} but one has to be careful since the infinite-ranged HK interaction will introduce non-local long-ranged entanglement between sites, which may invalid the assumption of these sophisticated tools (area-law of entanglement).

\section{Discussions}\label{sec4}
\subsection{Relations to other works}
Firstly, although the Ref.~\onlinecite{Leeb2023} begins with the HK model, their calculation is only performed in continuum limit since the projection into Landau level can be performed effectively for continuum models. In contrast, we focus on the lattice model and the lattice effect has been included, e.g. the Hofstadter butterfly spectrum in non-Fermi liquid state, quantum critical point and Mott insulator.

Secondly, Ref.~\onlinecite{Wasserman1996} seems to be the only review, which gives detailed treatment on the quantum oscillation with interaction. The main assumption of this review is that the self-energy has no strong momentum dependence, so one can neglect its magnetic field dependence when considering QO. Under this assumption, a generalized LK formula with imaginary-frequency-dependent self-energy can be derived. However, we know that the self-energy of HK model has explicit momentum dependence, e.g. at half-filling ($\mu=U/2$), $\Sigma(k,\omega)=\frac{(U/2)^2}{\omega-\varepsilon_{k}}$.\cite{Phillips2020} Therefore, the theory of Ref.~\onlinecite{Wasserman1996} cannot be used without non-trivial modification.

Finally, the work of Ref.~\onlinecite{Schlottmann2008} seems to be an interesting application of the theory in Ref.~\onlinecite{Wasserman1996}. However, due to the mentioned momentum dependence in self-energy, in our opinion, his results are not related to our model unless one can include the effect of momentum-dependence.

\subsection{Toward a theory for quantum oscillation in strongly correlated system}
As what has been discussed in last subsection, the widely-used generalized LK theory may not be useful if the self-energy has strong momentum-dependence. Currently, we have no ideas on how to bypass such difficulty. As emphasized in Ref.~\onlinecite{Chakravarty2011}, a system with self-energy like $\Sigma(k,\omega)\sim(\omega-v_{F}(k-k_{F}))^{\alpha}$ may be a good starting point. In addition, models solved by large-scale Monte Carlo simulation will provide new insights because NFL states, which violate Luttinger's theorem and exhibit strange metal behaviors, have been discovered.\cite{Yang2021b,Yang2022}

It is frank to say that after $60$ years, our understanding on QO in interacting systems is still based on the work of Luttinger.\cite{Luttinger1961} How to extend the framework of Luttinger into generic strongly correlated electron systems without Landau quasiparticle is an open question.\cite{Chakravarty2011,Denef2009}

\section{Conclusion and Future direction}\label{sec5}
In conclusion, we have taken Hatsugai-Kohmoto model as an example to study the quantum oscillation behavior in strongly correlated system. With Luttinger's approximation, it is found that although the non-Fermi liquid state has no Landau quasiparticle, the quantum oscillation indeed appears  and one can use Lifshitz-Kosevich-like formula to extract its basic properties. As a byproduct, Hofstadter butterfly exists in all phases of the ground-state whatever they are non-Fermi liquid or Mott insulator. We have also performed a small size exact diagonalization calculation, and its results exhibit certain oscillation behavior but we cannot extract exact value of oscillation period and amplitude. Therefore, we expect that future work on this interesting issue is highly desirable, particularly the large-scale numerical simulation.

\emph{Note added}: After completing this work, we have noticed the paper of Leeb and Knolle,\cite{Leeb2023} whose results generally agree with ours though their calculation is based on the projection into Landau level and works in continuum limit.

\appendix
\section{Derivation of singe-particle Green's function}\label{ap_A}
Follow the treatment of Hubbard model,\cite{Hubbard1963} let us define the single-particle Green's function as $G_{\sigma}(k,\omega)=\langle\langle \hat{c}_{k\sigma}|\hat{c}_{k\sigma}^{\dag}\rangle\rangle_{\omega}$, which is just the Fourier transformation of the retarded Green's function
\begin{equation}
G_{\sigma}(k,t)=-i\theta(t)\langle[\hat{c}_{k\sigma}(t),\hat{c}_{k\sigma}^{\dag}]_{+}\rangle.\nonumber
\end{equation}
Then, in terms of
\begin{eqnarray}
&&[\hat{c}_{k\sigma},\hat{H}]=(\varepsilon_{k}-\mu)\hat{c}_{k\sigma}+U\hat{c}_{k\sigma}\hat{n}_{k\bar{\sigma}},\nonumber\\
&&[\hat{c}_{k\sigma}\hat{n}_{k\bar{\sigma}},\hat{H}]=(\varepsilon_{k}-\mu+U)\hat{c}_{k\sigma}\hat{n}_{k\bar{\sigma}},\nonumber
\end{eqnarray}
we find
\begin{equation}
\omega\langle\langle \hat{c}_{k\sigma}|\hat{c}_{k\sigma}^{\dag}\rangle\rangle_{\omega}=1+(\varepsilon_{k}-\mu)\langle\langle \hat{c}_{k\sigma}|\hat{c}_{k\sigma}^{\dag}\rangle\rangle_{\omega}+U\langle\langle \hat{c}_{k\sigma}\hat{n}_{k\bar{\sigma}}|\hat{c}_{k\sigma}^{\dag}\rangle\rangle_{\omega}\nonumber
\end{equation}
and
\begin{equation}
\omega\langle\langle \hat{c}_{k\sigma}\hat{n}_{k\bar{\sigma}}|\hat{c}_{k\sigma}^{\dag}\rangle\rangle_{\omega}=\langle \hat{n}_{k\bar{\sigma}}\rangle+(\varepsilon_{k}-\mu+U)\langle\langle \hat{c}_{k\sigma}\hat{n}_{k\bar{\sigma}}|\hat{c}_{k\sigma}^{\dag}\rangle\rangle_{\omega}\nonumber
\end{equation}

Since above equations are closed, we obtain
\begin{equation}
\langle\langle \hat{c}_{k\sigma}\hat{n}_{k\bar{\sigma}}|\hat{c}_{k\sigma}^{\dag}\rangle\rangle_{\omega}=\frac{\langle \hat{n}_{k\bar{\sigma}}\rangle}{\omega-\varepsilon_{k}+\mu-U}\nonumber
\end{equation}
and
\begin{eqnarray}
G_{\sigma}(k,\omega)&=&\frac{1+\frac{U\langle\hat{n}_{k\bar{\sigma}}\rangle}{\omega-(\varepsilon_{k}-\mu+U)}}{\omega-(\varepsilon_{k}-\mu)}\nonumber\\
&=&\frac{1-\langle\hat{n}_{k\bar{\sigma}}\rangle}{\omega-(\varepsilon_{k}-\mu)}+\frac{\langle\hat{n}_{k\bar{\sigma}}\rangle}{\omega-(\varepsilon_{k}-\mu+U)}\nonumber
\end{eqnarray}
which is just the wanted Eq.~\ref{eq4} in the main text.

\end{document}